# UNDERSTANDING STUDENT AND ACADEMIC STAFF PERCEPTIONS OF AI USE IN ASSESSMENT AND FEEDBACK

Jasper Roe [1*], Mike Perkins [2], Daniel Ruelle [3]

[1] James Cook University Singapore, Singapore
[2] British University Vietnam, Vietnam
[3] VinUniversity, Vietnam

[*] Corresponding Author: jasper.roe@jcu.edu.au

June 2024

## Abstract

The rise of Artificial Intelligence (AI) and Generative Artificial Intelligence (GenAI) in higher education necessitates assessment reform. This study addresses a critical gap by exploring student and academic staff experiences with AI and GenAI tools, focusing on their familiarity and comfort with current and potential future applications in learning and assessment. An online survey collected data from 35 academic staff and 282 students across two universities in Vietnam and one in Singapore, examining GenAI familiarity, perceptions of its use in assessment marking and feedback, knowledge checking and participation, and experiences of GenAI text detection.

Descriptive statistics and reflexive thematic analysis revealed a generally low familiarity with GenAI among both groups. GenAI feedback was viewed negatively; however, it was viewed more positively when combined with instructor feedback. Academic staff were more accepting of GenAI text detection tools and grade adjustments based on detection results compared to students. Qualitative analysis identified three themes: unclear understanding of text detection tools, variability in experiences with GenAI detectors, and mixed feelings about GenAI's future impact on educational assessment. These findings have major implications regarding the development of policies and practices for GenAI-enabled assessment and feedback in higher education.

# Introduction

Generative AI (GenAI) refers to an Artificial Intelligence (AI) technique which can produce novel multimodal content, encompassing text, video, images, and more Luo (2024). The numerous release of publicly, and often freely accessible GenAI applications has resulted in their widespread adoption in both work and study environments, with OpenAI's ChatGPT having become the most rapidly expanding consumer application in history (Abdaljaleel et al., 2024).

In the educational domain, one of the most well-documented properties of GenAI is its potential to impact the academic integrity of assessments by enabling users to misrepresent authorship of written work, and this concern has been raised my multiple authors (Cotton et al., 2023; Perkins, 2023). Studies have identified how GenAI tools may tackle multiple forms of assessment with minimal human intervention. OpenAI's GPT-4 model (used in ChatGPT) performs comparably to human test-takers on MCQ formats across disciplines (Newton & Xiromeriti, 2024), and in a large-scale study across seven Australian universities, Nikolic et al. (2023) found that even slightly altered ChatGPT responses were sufficient to pass various assessments in multiple disciplines. Furthermore, research suggests that usage of these tools is prevalent; Chan (2023)'s findings demonstrate that up to a third of college students use GenAI in their assessed work, even despite believing that such use constitutes cheating or a violation of academic integrity rules. Further compounding this issue is the fact that at present, AI-created output cannot be accurately detected and common detection tools have highly variable efficacy (Chaka, 2024; Perkins et al., 2023; Weber-Wulff et al., 2023; Perkins, Roe, et al., 2024).

With the above in mind, assessment reform has come to the fore as a potential strategy for dealing with these risks while encouraging potentially legitimate and beneficial use cases of a new technology. Such reform strategies include directing student attention toward the skill of evaluative judgement (Bearman et al., 2024), focusing on open-ended assessments (Cotton et al., 2023), and using scalar approaches which enable the inclusion of GenAI where appropriate (Furze et al., 2024; Perkins, Furze, et al., 2024). GenAI applications may also be used to develop assessments and provide feedback. It has been suggested that teachers can use tools like ChatGPT to generate prompts for open-ended questions and develop rubrics (Baidoo-Anu & Owusu Ansah, 2023), and for scoring student work (Swiecki et al., 2022), providing feedback (Crawford et al., 2023; Dai et al., 2023), or wholly automating the marking process (Mizumoto & Eguchi, 2023; Ramesh & Sanampudi, 2022). However, critics of these approaches point to privacy and data risks (Nguyen et al., 2023) and the shifting of assessment responsibility to the developers of GenAI tools rather than the educator (Swiecki et al., 2022).

To make such assessment changes, and address both the potential risks and benefits of GenAI to assessment and feedback, it is necessary to understand and explore the views and experiences of students and academic staff. Yet there remains a gap in empirical studies addressing this topic, and focus has primarily been on the theoretical applications of GenAI, rather than on user experience and acceptance. This study aims to fill this gap by exploring how familiar students and academic staff are with AI and GenAI, and their comfort with the technology's use in both assessment and feedback, as well as understand their current experiences with this new technology. We choose to focus on three key areas which we feel reflect current and potential future uses of AI systems in assessment.

The first area is current familiarity – in other words, what is the current level of knowledge and awareness of these tools. Secondly, we assess how comfortable participants are with commonly posited use-cases of GenAI in assessment. In this area, we focus on feedback and marking on assessment items by AI alone and as an adjunct to traditional human feedback. We also explore the uses of AI systems in formative assessment including knowledge and participation checking. Finally, we focus on GenAI text detection, which represents one of the most recent GenAI-related technologies to become rapidly embedded in many Higher Education Institutions' (HEIs) assessment practice. A unique aspect of this study is the Southeast Asian context, as the research takes place across Singaporean and Vietnamese universities – a geographical location which is understudied in this topic. The results of this research contribute to the theoretical understanding of GenAI adoption in education while also providing insights that can inform policy and practice, ensuring that the implementation of new technologies incorporates voices from students and academic staff alike.

## Literature Review

There is a growing body of research exploring attitudes toward AI and GenAI among stakeholders in Higher Education (HE), and research has shown a pattern of optimism among HE students. Polyportis and Pahos (2024) studied 355 students in the Netherlands, and Vo and Nguyen (2024) surveyed 369 students in Central Vietnam, finding that Dutch students believed ChatGPT could improve academic performance, and Vietnamese students held a positive attitude towards the technology. Zhou et al. (2024) found that 28 entrepreneurship students reported that GenAI tools enhanced the learning experience and productivity. Additionally, Zhang et al. (2024) reported that over 80% of 850 Chinese university students were cautiously optimistic about using ChatGPT in education, and Wang et al.(2022) similarly found optimism in the possibilities of GenAI to support international students who often struggle with linguistic and stylistic challenges in written assessments. Among educators, results have been more mixed. Chan and Lee (2023) and Sevnarayan and Potter (2024) found that teachers tended to be more sceptical of GenAI's capabilities compared to students and are more concerned about its impact on academic integrity in assessments.

In terms of willingness to use GenAI in learning, there are mixed findings. Chan and Zhou (2023) found a positive correlation between perceived value and intention to use GenAI among 405 students, with a weak negative correlation between perceived cost and intention to use. Kim et al. (2020) examined perceptions of AI among higher education teaching assistants, finding stronger support for their adoption than academic staff. Smolansky et al. (2023) found that educators were more concerned about the impact of GenAI in assessments than students, with greater vulnerability in written text assignments (essays, case studies, reports), computer code, and quiz questions (multiple choice, short answer). Interestingly, they found that students' reactions to GenAI in education were more optimistic, particularly in the boosts of productivity that GenAI offers, despite some concerns about potential losses of creativity and personalized feedback.

Regarding attitudes toward assessment practices and academic integrity, Firat (2023) interviewed seven academics and 14 PhD students from multiple countries, with a consensus that GenAI stands to significantly transform assessment practices. Lee et al. (2024) surveyed 30 teaching staff at an Australian university, noting a lack of coherent views on GenAI in higher education, though nearly half were already using GenAI in their roles, primarily related to assessment practices. The study also highlighted the inadequacy of AI-text detection tools in assessments, aligning with the findings of Perkins et al. (2024) and Weber-Wulff et al. (2023), both of which discuss the inadequacy of detection tools to accurately determine whether academic integrity violations have occurred. Ghimiere et al.(2024) found a consensus among 116 educators that GenAI tools will become integral to education, with benefits seemingly outweighing the negatives, but concerns from both educators and students about the impact of GenAI on assessment and academic integrity are a recurring theme.

Xia et al. (2024) conducted a scoping review of research on GenAI in assessment, developing recommendations to transform assessment practice, highlighting opportunities for building digital literacy among teaching staff and encouraging innovation in assessment and teaching practices. Challenges related to academic integrity and distrust of detection tools remain, with Luo (2024) conducting an in-depth study with 11 education students, revealing that high AI scores from Turnitin led to significant adverse consequences and that the trust in professors was linked to AI use and declarations. AI text detection in assessed work as a method of safeguarding academic integrity has received significant attention and a consensus is emerging that these tools do not demonstrate enough accuracy to be used as a standalone method for identifying GenAI-produced text (Perkins et al., 2023; Perkins, Roe, et al., 2024; Weber-Wulff et al., 2023). Furthermore, they may demonstrate biases against certain student groups, threatening equity of assessment (Liang et al., 2023).

**Research Questions**

Based on the gaps in the present literature, our aim for this research is to understand familiarity and comfort of students and academic staff with GenAI in assessment practices, as well as to explore current experiences with a recent technology – GenAI text detection tools. Our research questions for this study are as follows:

1. How familiar are academic staff and students with GenAI tools?
2. How comfortable are academic staff and students with GenAI tools in assessment practices?
3. How comfortable are academic staff and students with GenAI text detection?
4. What are academic staff and student's experiences of GenAI text detection?

# Methodology

After receiving ethical approval from an internal review board we developed a questionnaire to be delivered to academic staff and students studying at three university campuses across Southeast Asia (two in Vietnam, one in Singapore) combining both quantitative Likert-type scale questions with open-ended text entry questions. The survey combined quantitative Likert-type scale questions with open-ended text entry questions.

**Participants and Sampling Strategy**

We employed a multi-channel convenience sampling approach to maximize reach and representation across our target population. The survey was distributed through institutional newsletters, email lists, in-class promotions, and campus posters, aimed at capturing a cross-section of students and academic staff across all disciplines and study levels. All participants were above 18 years old and included students from multiple study levels, including pre-university pathway programs, undergraduate and postgraduate study.

**Data Collection**

Like Spooren et al. (2007), we adopted the Likert-type scale for its ease of use and straightforward nature, making it applicable to this form of educational research. However, we note that such scales are highly variable (Roszkowski & Soven, 2010) which means that comparison of results across studies is challenging. In this survey, our initial question used a 5-point scale to determine degrees of familiarity (ranging from very familiar to unfamiliar), which represents a common approach. Subsequent questions used a standard 7-point scale ranging from strongly disagree to strongly agree, (1 – 7) and included neutral options. We decided to include neutral options as AI technologies are new and complex in nature, thus we expected a significant degree of uncertainty among respondents and consider this important to understand. At the same time, we captured qualitative data using an open-ended question, in order to give participants control over the thoughts they were able to share and the language used to share them (Scotland, 2016). The survey was hosted on the Qualtrics platform which allowed for advanced question branching.

**Data Analysis**

Similar studies focused on an exploratory approach to an assessment phenomenon have used a combination of descriptive statistics with in-depth qualitative analysis fruitfully (Stančić 2021), and so we followed a similar method. Our quantitative questions were analysed using descriptive statistics while we employed reflexive thematic analysis (TA) (Braun et al., 2022; Braun & Clarke, 2006, 2019) for the qualitative data. In this case, the reflexive element means that we interpreted the data through our own lens of experience as educators, centrally recognising our own subjectivity and seeking to develop a sense of meaning from our responses, rather than search for a positivistic sense of 'truth' in answering our research questions. As a result, our reflexive analysis of the data led the output of codes, subthemes, and overarching themes; these outputs were arrived at through the creative labour of the coding process (Braun & Clarke, 2019) and thus represent the reflexive TA approach, as opposed to the more common topic summaries which may also be labelled as TA (Braun et al., 2022). This method was structured around the common six-step process for thematic analysis as described by Braun & Clarke (2006), including data familiarization, generation of initial codes, searching for themes, reviewing themes iteratively, and then defining and naming themes, prior to producing a final report.

The questions used in this survey can be seen in Table 1:

| # | Question List (Academic staff) | Question List (Student) |
|---|---|---|
| 1 | Generative AI is defined as artificial intelligence capable of generating text, images, or other media. How familiar are you with Generative AI (GenAI)? | Generative AI is defined as artificial intelligence capable of generating text, images, or other media. How familiar are you with Generative AI (GenAI)? |
| 2 | I believe that an AI system marking student essays or examinations would grade students fairly. | I believe that an AI system marking student essays or examinations would grade students fairly. |
| 3 | I would feel comfortable with an AI system giving students immediate feedback on their written work, without any other feedback from a teacher. | I would feel comfortable with an AI system giving me immediate feedback on my written work, without any other feedback from a teacher. |
| 4 | I would feel comfortable with an AI system giving students feedback on their written work, as long as they had feedback from a teacher later. | I would feel comfortable with an AI system giving me feedback on my written work, as long as I had feedback from a teacher later. |
| 5 | I would feel comfortable with my university using an AI system to check whether students have understood their course material and suggest additional resources if they do not. | I would feel comfortable with my university using an AI system to check whether students have understood their course material and suggest additional resources if they do not. |
| 6 | I would feel comfortable with an AI system analysing how much students participate in online classes. | I would feel comfortable with an AI system analysing how much students participate in online classes. |
| **AI Detection Questions** | | |
| 7 | I would feel comfortable with using a GenAI detector to see if students have used AI in their written work. | I would feel comfortable with teachers using a GenAI detector to see if students have used AI in their written work. |
| 8 | It is fair for a teacher to lower students' grades based on what an AI text detector says. | It is fair for a teacher to lower students' grades based on what an AI text detector says |
| **Open Ended Questions** | | |
| 9 | Have you ever used an AI text detector on student work? Please tell us about your experience. | Have you ever had a teacher use an AI text detector on your work? Please tell us about your experience. |
| 10 | Do you have any other information you would like to share regarding your experiences of GenAI and/or AI text detection in the classroom? | Do you have any other information you would like to share regarding your experiences of GenAI and/or AI text detection in the classroom? |

*Table 1 Survey Questions*

## Results

**Descriptive statistics**

Our questionnaire received 35 responses from university academic staff and 282 responses from students, resulting in 317 total responses, split approximately equal across the three institutions. However, 48 students did not complete questions 3 – 8 while 1 academic staff respondent did not complete questions 4 – 8. These partial responses were included in the analysis for questions 1 - 2 but were excluded from descriptive statistics calculations for questions 3 – 8 and 4 – 8 respectively. The descriptive results for both student and academic staff responses are shown in Table 2.

| Question | Group | N | Mean | Median | Mode | Std. Deviation |
|---|---|---|---|---|---|---|
| 1. Familiarity with GenAI | Academic Staff | 35 | 1.914 | 2.000 | 2.000 | 0.887 |
| | Students | 282 | 1.826 | 2.000 | 2.000 | 0.770 |
| 2. AI marks Fairly | Academic Staff | 35 | 3.429 | 3.000 | 2.000 | 1.539 |
| | Students | 282 | 3.309 | 3.000 | 3.000 | 1.505 |
| 3. AI can give feedback independently | Academic Staff | 35 | 2.857 | 2.000 | 2.000 | 1.498 |
| | Students | 235 | 3.179 | 3.000 | 2.000 | 1.623 |
| 4. AI can give feedback with teacher feedback | Academic Staff | 34 | 4.912 | 6.000 | 6.000 | 1.832 |
| | Students | 234 | 5.316 | 6.000 | 6.000 | 1.640 |
| 5. AI can check student understanding | Academic Staff | 34 | 5.324 | 5.000 | 5.000 | 1.121 |
| | Students | 234 | 4.987 | 5.000 | 6.000 | 1.587 |
| 6. Comfort with AI participation analysis | Academic Staff | 34 | 5.971 | 6.000 | 6.000 | 1.314 |
| | Students | 235 | 4.740 | 5.000 | 6.000 | 1.676 |
| 7. Comfort with the use of GenAI text detectors | Academic Staff | 34 | 5.382 | 6.000 | 6.000 | 1.907 |
| | Students | 235 | 4.885 | 5.000 | 6.000 | 1.722 |
| 8. Comfort with the use of AI detection results to lower grades | Academic Staff | 34 | 3.882 | 4.500 | 5.000 | 1.822 |
| | Students | 235 | 3.191 | 3.000 | 2.000 | 1.612 |

*Table 2 Descriptive statistics*

Our analysis of survey responses reveals varying perceptions of GenAI across four key areas in educational assessment. The following sections present descriptive statistics for each category, highlighting differences between academic staff and student views.

*Familiarity*

Both academic staff and student respondents demonstrated low familiarity with GenAI, as evidenced by low means (academic staff: 1.914, students: 1.826) and low standard deviations (academic staff: 0.887, students: 0.770), indicating a relatively uniform lack of familiarity.

*Assessment Marking and Feedback*

There were varied opinions on using GenAI for marking assessments. Academic staff had a mean of 3.429 and median of 3, with a high standard deviation of 1.539, indicating diverse opinions. Students were moderately sceptical, with a mean of 3.309 and median of 3. When considering GenAI's capability to provide independent feedback, students were slightly more comfortable (mean 3.179, median 3, SD 1.623) compared to academic staff (mean 2.857, median 2, SD 1.498). Both groups were more comfortable with GenAI providing feedback when combined with instructor feedback (academic staff mean 4.912, median 6, SD 1.832; students mean 5.316, median 6, SD 1.640).

*Knowledge Checking and Participation*

Academic staff expressed comfort with GenAI systems checking student understanding (mean 5.324, median 5, SD 1.121) and analysing online participation (mean 5.971, median 6, SD 1.314). Students also showed reasonable comfort with GenAI for these tasks (checking understanding: mean 4.987, median 5, SD 1.587; analysing participation: mean 4.740, median 5, SD 1.676).

*GenAI Text Detection*

Academic staff were relatively comfortable with using GenAI detectors (mean 5.382, median 6, SD 1.907), but were more divided on lowering grades based on detection results (mean 3.882, median 4.5, SD 1.822). Students showed a slightly lower level of comfort with GenAI detection (mean 4.885, median 5, SD 1.722) and were generally opposed to lowering grades based on detection results (mean 3.191, median 3, SD 1.612).

**Thematic Analysis**

We collected a combined 33 responses to Question 9 & 10 from academic staff, and 121 responses from students. To build a holistic picture of GenAI and text detection in learning and assessment, we decided to group the data together for analysis (i.e. not analysing separately by category of academic staff or student). We developed three overall themes from the data, visible in Table 3:

| Themes | Sub-Themes | Initial Codes |
|---|---|---|
| Qualified understanding of policies and technologies. | Unclear understanding of how AI text detection tools work. | • Detection focuses on ideas not words.<br>• Belief that detection is used but unsure of details.<br>• Belief detection is used for screening only.<br>• Conflating plagiarism detection with AI detection. |
| | Unclear understanding of university policies. | • Belief that any use of GenAI is prohibited.<br>• Belief that GenAI is not allowed for teaching.<br>• Unsure/unclear if AI detection is used. |
| Variable Detection Results. | False positives, false negatives and false accusations. | • Frustrating experiences of false positives.<br>• Unfair thresholds for detection.<br>• Inaccurate results returned. |
| | Semi-accurate usage. | • Tool clunky and hard to use.<br>• Used as a starting point for discussion.<br>• Usually accurate but with occasional false positives. |
| Mixed Feelings for the Future. | Perceived benefits of AI use. | • AI as a helper with writing.<br>• AI as a resource for questions and answers.<br>• Students need to be trained. |
| | Fears over impact. | • Role of teachers and students will change.<br>• Loss of authenticity of content.<br>• Homogeneity of knowledge. |

*Table 3 Identified themes, sub-themes and codes*

*Qualified Understanding of Policies and Technologies*

While many respondents indicated their comfort with the use of detection technology, a recurring element was a lack of certainty on exactly what part of an assessment was being examined by GenAI detection technologies and why. Most often, this theme was reflected in simple answers of 'unsure' or 'not sure' when responding to Question 9. However, in Question 10, several narratives elaborated on instances of text detection which seemed to misinterpret what detection tools are capable of. One such example is below:

> *"The result came back was 0% because the ideas are original. However, I admit that I used ChatGPT to improve my writing and correct my grammar."*

This response to Q10 suggests that the respondent (a student) believed that it is the originality of the idea rather than the structure of the response which would lead to a positive detection result for an AI detection tool when GenAI text detection in reality relies on linguistic and structural features. A qualified understanding therefore describes a limited awareness of not only whether detection technologies are used, but how they operate and their abilities. Further to this, responses also commonly indicated a degree of confusion on institutional policies. In some specific cases, students provided views that GenAI was completely disallowed by an institution, which runs counter to the actual publicly available institutional policies. Overall, this suggests that confusion and a lack of clarity over AI and its capabilities and permissibility does not just relate to the technology, but also to the understanding of policy in higher education.

*Variable Detection Results*

This theme captures the range of different experiences relating to the use of AI and AI text detection. Many respondents shared narratives in which they have been falsely accused of using AI, or have known others who have experienced accusations, while academic staff also relayed cases in which they had attempted to use GenAI text detection and encountered false positives. No responses completely endorsed the use of AI text detectors, although some academic staff gave a nuanced account of their validity and need to develop into the future:

> *"I believe the current AI detection software(s) being utilized at my university does a "fair" job of detecting blatant plagiarism or AI usage for student assignments, but it clearly is not a catch-all* *solution with 100% detection rates. As students become increasingly familiar with AI and ways to circumvent AI detection models, I feel the complexity of AI detectors will need to improve in order to remain relevant."*

In contrast, other responses from students gave detailed accounts of the inadequacies of AI text detection software, leading to leading to perceived major negative impacts. One respondent phrases an encounter as follows:

*"It was terrible. I am a student with consistently above average to good grades, and I also actively participate in classes. Despite this, when an assignment of mine was flagged as "highly likely to have used AI", the subject coordinator against the advice of BOTH my lecturer and tutor had me REDO a 2000 word assignment, which I dedicated a lot of time to complete. It is true that AI was used in the writing of my assignment, however it does not equate to it being written by AI. I specifically used it for advice and structural purposes; not in the composition of unique and critical ideas. As a first language speaker and a decent student, this experience was a huge stain on my academic life. It was terrible and I hope nobody else would have to experience what I did."*

This case highlights the potentially severe consequences of a false accusation. Regardless of the veracity of the response, it becomes clear that to avoid situations in which an assessor and student have different perspectives on the acceptability of AI, clear, mutual guidelines need to be established. At times, the justification for lowering assessment grades can be made based on questionable outputs, as in the following:

*"Yes, a TA lowered our grade by saying that it was created by AI, we were surprised and made an appeal. It actually turned out that he based on 16% likelihood to be created by AI and failed our questions without even reading the content. It was extremely annoying. Eventually, after further analysis with us, he agreed to give us our deserved marks."*

Of these responses, a recurring subtheme was frustration and inequity – learners reported feeling that their assessors had not taken a nuanced view of the perceived use of AI, and rather had enacted consequences that felt out of proportion to the work that they had submitted. At times, even accusations without 'judgement' led to disruptions for students:

*"I think a teacher used an AI text detector on an essay I wrote and suspected it was written by AI, which it wasn't, causing the grading of the assignment to be delayed. This was quite frustrating, especially since the teacher gave no other input on the matter."*

Among academic staff respondents, similar cases of false positives and false negatives were reported, even among those who indicated positive inclinations towards the use of such tools:

*"AI detection is usually accurate but it did generate some misleading results and we had to call the student in for a viva, and the student passed it."*

Other academic staff responses discussed cases of testing out detectors using AI generated content, and identifying false negatives:

*"I tried <Name of AI detector> and it did not work well. Multiple examples of ChatGPT content I fed it were returned as having no AI input."*

Overall, these findings contrast with the results of the descriptive statistics, in which most respondents agreed with the use of GenAI text detectors, and many also agreed that they could be used to justify the lowering of grades.

*Mixed Feelings for the Future*

A final theme that emerged from the data relates to the qualitative data was that of a range of views on how the development of AI will affect assessment and education in the future.

In some student responses, there was a noted desire to use GenAI autonomously for structural improvements to writing assessments, broadly in line with current policies surrounding ethical use of GenAI tools (Perkins & Roe, 2023a, 2023b).

*"As long as the ideas are mine, structure of answers are mine, the usage of AI will be beneficial. AI will be extremely helpful in correcting your grammars, rewriting your answers more clearly (still based on your understanding)."*

> *"I could ask Bard (google) questions with my literature review or how I have phrased it wrongly and it could give me immediate corrections, especially in detail. Which I may not be comfortable asking lecturers at times, in case it is a stupid question. Or need explanations at length."*

Academic staff responses equally conceptualised AI as a morally neutral tool, in which teachers played an important role in providing critical AI literacy education:

> *"GenAI functions as a tool, both presently and in the future. The crucial factor lies in how we wield this tool. Teachers play a pivotal role in instructing students on utilizing GenAI to accomplish tasks effectively, while also fostering the ability to discern and address its limitations."*

In contrast to these hopeful and beneficial implications of GenAI on assessment, learning, and teaching, other responses highlighted potential risks and impacts on the educational process. This at times related to the ideas of a monoculture of knowledge, an implication of AI development outlined by Messeri & Crockett (2024). Under this scenario, the respondent considers whether using GenAI will eventually result in unwanted lack of distinctiveness in programs of higher education:

> *"There is a risk of homogeneity in content, from both teachers and students, with GenAI, even with well written and curated prompts. What's the USP of a course once it is heavily tooled with GenAI?"*

While similarly, for one academic staff respondent, this potentially leads to forms of educational assessment in which AI evaluates AI, thus removing any 'human' component from the task:

> *"I have felt AI usage is killing the critical thinking abilities of students. Also, by allowing AI usage, I have felt me as a human evaluating an AI generated content."*

# Discussion

The study revealed both academic staff and student scepticism towards using GenAI systems to mark assessments, with significant variation in acceptability. However, both groups appeared more comfortable with AI generated feedback when it supplemented traditional, teacher-delivered feedback. This aligns with findings from Chan and Lee (2023) and Sevnarayan and Potter (2024), who noted that teachers tend to be more sceptical of GenAI's capabilities compared to students. The implication of these results is that educational institutions could consider piloting such hybrid models to enhance assessment practices. However, we did not specify what such an AI system would entail, indicating a need for further research into different methods of generating and integrating AI feedback across various assessment types. Future studies could explore specific AI integration models and their effectiveness in different educational contexts.

Both academic staff and students generally accepted the use of AI systems for checking knowledge and monitoring student participation. This acceptance aligns with the findings of Kim et al. (2020), who found stronger support for AI adoption among teaching assistants compared to academic staff. This highlights an area for further exploration, particularly given that current educational technologies, like Virtual Learning Environments (VLEs), already use analytics data for these purposes. However, the variability in responses suggests that any implementation of these systems must involve clear communication about the degree of surveillance and the nature of the data collected. Consequently, future research could examine the ethical implications of AI surveillance in education and develop guidelines to ensure transparency and student consent, addressing concerns raised by Nguyen et al. (2023) regarding privacy and data risks.

Despite the overall comfort with GenAI text detectors, qualitative responses revealed several concerns. Students and academic staff recounted instances of false positives and misinterpretations of AI detection results, sometimes leading to severe consequences for students and causing confusion for academic staff. These narratives highlight ethical and equity issues related to the use of AI detection tools, echoing the findings of Perkins et al. (2023) and Weber-Wulff et al. (2023) regarding the inadequacy of detection tools. This suggests that while these tools are broadly acceptable, their application requires careful consideration and refinement. As a result, developers of AI detection tools need to improve their accuracy and provide clearer guidelines for their use. Future research could focus on refining AI algorithms to reduce false positives and on developing comprehensive training programs for users.

The theme of Qualified Understanding and Confusion reflects a lack of clarity about how detection technologies work and their applications. This confusion extends to both students and academic staff, emphasizing the need for clear, coherent frameworks to guide the use of AI in higher education. Consequently, educational institutions should invest in training programs that enhance understanding and confidence in AI tools in assessment. Future studies could investigate the impact of such training programs on the acceptance and effectiveness of AI in education.

Overall, our findings suggest mixed feelings about the role of AI in assessments and higher education, echoing the varied perspectives found by Ghimiere et al. (2024). The variability in responses indicates strong opinions both for and against AI use. Some view AI as a valuable aid in learning and assessment, useful for providing answers, improving writing, and handling routine tasks. However, there are concerns about the potential for knowledge loss, content homogenization, and the loss of unique educational experiences. Academic staff expressed these fears more frequently, while students showed more enthusiasm for engaging with AI tools, aligning with the findings of Zhang et al. (2024). The implication is that any implementation of AI in educational assessment must balance these concerns by ensuring that AI complements rather than replaces traditional educational methods. Future research could explore strategies for integrating AI in a way that preserves the unique aspects of educational experiences while leveraging the benefits of technology.

## Conclusion

This study offers critical insights into the perceptions of academic staff and students regarding GenAI in assessment in higher education. Our research questions focused on familiarity with GenAI tools, comfort with their use in assessment practices, attitudes towards GenAI text detection, and experiences with these detection tools. Regarding familiarity (RQ1), we found that both academic staff and students generally demonstrated low familiarity with GenAI tools. In terms of comfort with GenAI in assessment practices (RQ2), there was skepticism towards AI-only marking but greater acceptance of AI-assisted feedback when combined with traditional teacher input. Both groups showed acceptance of AI for knowledge checking and participation monitoring. For GenAI text detection (RQ3), we observed a general comfort with its use, though this was tempered by concerns about accuracy and fairness. Experiences with GenAI text detection (RQ4) varied widely, with reports of both false positives and negatives, highlighting the need for careful implementation and clear communication about these tools.

Although we draw on a small sample size and do not differentiate by institution, academic discipline or stage of study, a holistic interpretation of results reveals relevant patterns and trends regarding AI and assessment. These findings provide a foundation for developing policies and strategies for the effective use of GenAI in educational assessment and feedback. Consequently, greater research across geographies and cultural contexts is necessary to complement our findings. These findings offer a preliminary insight into an important topic in the future of assessment of learning in an AI-enabled higher education context.

**AI Usage Disclaimer**

This study used Generative AI tools for revision and editorial purposes throughout the production of the manuscript. Models used were Claude 3 (Opus) and Claude 3.5 (Sonnet). The authors reviewed, edited, and take responsibility for all outputs of the tools used in this study